\documentclass[]{pasj02} 

\usepackage{natbib} 

\jyear{2025}
\Received{}
\Accepted{}

\graphicspath{{./}{figures/}} 

\begin{document} 

\title{A new approach for predicting the stability of hierarchical triple systems -- I. Coplanar Cases}

\author{
 Ryutaro \textsc{Iwakura},\altaffilmark{1}\footnotemark[*]\orcid{0009-0009-8049-2968} \email{riwakura@stu.kobe-u.ac.jp} 
 Yoko \textsc{Funato},\altaffilmark{2}\orcid{0000-0002-6992-7600} 
 Junichiro \textsc{Makino},\altaffilmark{1}\orcid{0000-0002-0411-4297} 
\altaffiltext{1}{Department of Planetology, Graduate school of Science, Kobe University, 1-1 Rokkodai-cho, Nada-ku, Kobe, Hyogo 657-8501, Japan}
\altaffiltext{2}{General Systems, Graduate School of Arts and Science, The University of Tokyo, Komaba, Meguro, Tokyo, 153-8902, Japan}
}



\maketitle

\begin{abstract}
Hierarchical triple systems play a crucial role in various astrophysical contexts, and therefore the understanding of their stability is important.
Traditional empirical stability criteria rely on a threshold value of $Q$, the ratio between the outer orbit's pericenter distance and the inner orbit's semi-major axis.
However, determining a single critical value of $Q$ is impossible because there is a range of the value of $Q$ for which both stable and unstable systems exist, referred to as the mixed-region.
In this study, we introduce a novel method to assess the stability of triple systems within this mixed-region.
We numerically integrate equal-mass, coplanar hierarchical triples within the mixed-region.
By performing Fourier analysis of the time evolution of the semi-major axes ratio during the first 1000 inner orbital periods of the systems, we find notable features in stable systems: 
if the main peaks are periodically spaced in the frequency domain and the continuous components and irregularly spaced peaks are small, the system tends to be stable.
This observation indicates that the evolution of stable triples is more periodic than that of unstable ones.
We quantified the periodicity of the triples and investigated the correlation between the Fourier power distribution and the system's lifetime.
Using this correlation, we show that it is possible to determine if a triple system in the mixed-region is stable or not with very high accuracy.
These findings suggest that periodicity in orbital evolution can serve as a robust indicator of stability for hierarchical triples.

\end{abstract}


\section{Introduction}
A hierarchical three-body system consists of a binary and a third body orbiting around the binary.
The stability of hierarchical three-body systems is a long-standing problem in the fields of dynamical astronomy, classical mechanics, and mathematics.
Understanding its stability is not only of theoretical interest but also a practical necessity for the interpretation of astronomical phenomena.

For example, a highly accurate three-body stability condition is essential for $N$-body simulations of globular clusters \citep{Aarseth2003, HH2003}.
In $N$-body simulations, the computational cost of binary and hierarchical three-body systems is very high, because their orbital timescale is many orders of magnitudes shorter than the dynamical or thermal timescale of the parent cluster.
Accordingly, in typical $N$-body simulations, the evolution of binary systems is computed in the following manner.
In the case of stable systems, we can integrate them using some variation of orbit averaged perturbation methods such as the slow-down method \citep[e.g.,][]{MA96, sdar}.
These methods are designed to reduce the number of time steps needed for orbit calculations while maintaining high accuracy.
On the other hand, for unstable systems, we cannot use such approximate methods because such systems should be integrated without approximation.
This is because applying approximation to unstable systems can lead to physically incorrect solutions.
In principle, when integrating such unstable systems without approximation, the computational cost would not be very large, since such unstable systems disintegrate in relatively short timescales.
Thus, prediction of the stability of triple systems with high accuracy is essential for efficient and accurate simulation of globular clusters.

However, there is no such high-accuracy prediction method.
A common issue is the misclassification of the dynamical stability of these systems --- that is, mistaking a stable system as an unstable one or vice versa.
Such misidentification can severely degrade computational efficiency by causing unnecessary direct integrations, effectively stalling the simulation.
Moreover, incorrect classification may result in physically inaccurate outcomes.

Here, we provide a brief review of hierarchical triple stability criteria.
Traditionally, numerous stability conditions have been formulated as expressions that denote the minimum value of $Q$:
\begin{equation}
    Q = \frac{q_\mathrm{out}}{a_\mathrm{in}} = \frac{a_\mathrm{out}(1-e_\mathrm{out})}{a_\mathrm{in}},
    \label{eq:q}
\end{equation}
where $q$ is the pericenter distance, $a$ is the semi-major axis, $e$ is the eccentricity, and subscripts ``in'' and ``out'' correspond to ``inner'' and ``outer'' orbit, respectively.
We define the critical value of the stability parameter $Q$, denoted as $Q_\mathrm{crit}$, as the minimum value above which the triple system remains dynamically stable.
Using semi-analytical or numerical methods, $Q_\mathrm{crit}$ is fitted as a function of the orbital elements of the system, i.e., $Q_\mathrm{crit} = Q_\mathrm{crit}(m_1, m_2, m_3, a_\mathrm{in}, a_\mathrm{out}, e_\mathrm{in}, e_\mathrm{out}, I, \dots)$,
where the variables include the component masses, semi-major axes, eccentricities, and orbital inclinations.
Dependence on other parameters such as the arguments of periapsis ($\omega$) or orbital phases is usually neglected.
This is because including those parameters would make the parameter survey significantly more computationally expensive, and the resulting criterion would become less practical for application.
By comparing the actual value of $Q$ for a given three-body system with the fitted $Q_\mathrm{crit}$, one hopes to judge the system to be unstable if $Q < Q_\mathrm{crit}$, and stable if $Q > Q_\mathrm{crit}$.

\citet{Harr72} found that the minimum value of $Q$ for stability differs between prograde and retrograde orbits.
Harrington also derived a stability criterion for coplanar orbits including three-body mass dependency \citep{Harr75, Harr77}.
\citet[hereafter EK95]{EK95} introduced a parameter $Y$, which is defined as the ratio of the apocenter distance of the inner to the pericenter distance of the outer, instead of using $Q$.
They also empirically derived a stability criterion in terms of $Y$.
Both $Q$ and $Y$ represent the distance between the inner and outer binaries, with no substantial difference between them.
\citet[hereafter MA01]{MA99, MA01} proposed a well-known dynamical stability criterion for hierarchical triples:
\begin{equation}
    Q_\mathrm{crit, MA01} = 2.8 \biggl ( 1-\frac{0.3I}{\pi}\biggr )\biggl [ \biggl (1+\frac{m_3}{m_1+m_2}\biggr ) \frac{1+e_\mathrm{out}}{\sqrt{1-e_\mathrm{out}}}\biggr] ^{2/5},
    \label{eq:MA01}
\end{equation}
where $I$ is the inclination, $m$ is the mass and $e$ is the eccentricity.
\citet[hereafter M18]{M18} further modified the MA01 criterion by incorporating the dependence on the inner eccentricity, while \citet[hereafter V22]{Vynatheya22} incorporated the effect of ZLK mechanism \citep{Zeipel, Lidov, Kozai} and refined the criteria proposed by EK95 and MA01.
Numerous three-body stability conditions have been reported for use in $N$-body simulations \citep[e.g.,][]{VMOR, Georgakarakos13, MK20, Hayashi+22}. 
Recently, several studies using machine learning have been reported \citep[e.g., V22; ][hereafter LT22]{LT22}.

Any formula for $Q$ proposed so far does not really discriminate stable and unstable systems, since for the values of $Q$ close to the ``critical''  value of $Q$, $Q_\mathrm{crit}$, for any of these criteria, some realizations of the three-body systems turned out to be stable while some to be unstable.
In this parameter space, both stable and unstable systems coexist, forming what is commonly referred to as a mixed-region \citep[e.g.,][]{Dvorak1986}.
One reason for the existence of this mixed-region is that in most formulas orbital elements such as the argument of periapsis $\omega$ or the initial phase are not taken into account, although these elements can significantly influence the stability of three-body systems \citep{Hayashi+22, Hayashi+23}.
Even if one attempts to extend the fitting procedure to include these additional orbital elements, the intrinsic chaotic nature of the three-body problem makes precise classification difficult.
This is particularly true in the mixed-region, where the system exhibits strong chaos, and small differences in orbital elements can lead to large variations in the lifetime of the system.
Conventional criteria based on $Q$ are designed to predict stability solely from initial conditions, which inherently limits their ability to account for chaos.

Therefore, to achieve high-accuracy stability classification in the mixed region, it is essential to evaluate not only the initial parameters but also the subsequent orbital evolution itself.
LT22 simulates triple systems with Q values 5\%–15\% below the MA01 critical threshold and employs machine learning on the orbital elements obtained from the simulations, achieving an AUC of about 0.95.
However, their approach uses a relatively large time step for sampling, and it remains unclear which aspects of the orbital elements are associated with instability.
In addition, the AUC value of 0.95 is rather low to be used in $N$-body simulations.
Hence, in order to evaluate stability by taking into account the orbital evolution, a more detailed analysis of the dynamical behavior of orbital elements is required.

The aim of this study is to develop a new method to determine the stability of hierarchical triples that can overcome the problems discussed above.
First, we conducted numerical integrations on coplanar equal-mass hierarchical three-body systems whose $Q$ at the initial condition is in the mixed-region.
Here we adopt the critical value $Q$ by MA01 as $Q_\mathrm{crit}$ of the mixed-region.
Then, we investigate the differences between long-lived and short-lived systems from the perspective of orbital element periodicity.

Our main results are the following:
First, with $Q$ value smaller than the critical value of MA01 criterion, most triples disintegrate by the time between $10^5P_\mathrm{in}$ and $10^6P_\mathrm{in}$ while there are highly stable triples that survive beyond $10^9P_\mathrm{in}$, where $P_\mathrm{in}$ is the initial inner orbital period.
Even if integration starts from the same $Q$ value, differences in phase and argument of periapsis ($\omega$) lead to differences of several orders of magnitude in the survival time, reflecting the chaotic behavior of triples \citep{Hayashi+22, Hayashi+23}.
Therefore, the conventional approach of describing stability criteria using $Q$ do not capture the fundamental nature of the system.
Second, the result of Fourier analysis of the time evolution of orbital elements over the initial $10^3P_\mathrm{in}$ shows that stable triples exhibit much periodic orbital evolution compared to those of unstable ones.
When observed in the frequency space of the time variation of the orbital elements, systems with clearly defined peaks at the fundamental frequency and its integer multiples tend to be more stable.
Surprisingly, this difference in orbital periodicity is already evident in the initial $10^3P_\mathrm{in}$, an early stage of evolution, even though many ``unstable'' systems have lifetimes $\sim 10^6P_\mathrm{in}$.
This property enables the identification of a majority of highly stable triples.
As a result, we have achieved highly accurate prediction of stability with AUC as high as 0.99.
For further details, see the Discussion section.

The plan of this paper is as follows:
In Section 2, we describe our models and method for our numerical integration.
We show the result of our simulation and the performance of our new stability criterion in Section 3.
Section 4 is for discussion.

\section{Methods}\label{sec:methods}
In this section, we describe the initial conditions and numerical methods used in our study.
We numerically integrated hierarchical triple systems with $Q$ values smaller than the lower limit determined by the MA01 criterion until the hierarchical structure of the systems was disrupted or the time reached to $10^9P_\mathrm{in}$.
Only Newtonian gravity is considered, neglecting any general relativistic effects.

\subsection{Initial Conditions}
Our coplanar three-body system consists of a inner binary and a third body, which we refer to as "inner" and "outer".
The two bodies that form inner binary has indices 1 and 2, while the outer body has index 3.
We consider two types of coplanar cases, prograde orbits with inclination  $I = 0$ and retrograde orbits with $I = \pi$.

We fixed some of the parameters and initial orbital elements: mass ($m_1 = m_2 = m_3 = 0.5 M_\odot$), inner semi-major axis ($a_\mathrm{in} = 1$ au), eccentricity ($e_\mathrm{in} = 0.5, e_\mathrm{out} = 0.25$) and inner argument of periapsis ($\omega_\mathrm{in}=0$).
These parameters are not special values but rather arbitrary ones.
We uniformly sample every $2\pi/10$ radian of outer arguments of periapsis ($\omega_\mathrm{out} \in \mathcal{U}[0, 2\pi)$) and mean anomalies ($M_\mathrm{in}, M_\mathrm{out} \in \mathcal{U}[0, 2\pi)$).
The list of initial conditions is shown in Table \ref{tab:elements}.

\begin{table}[htbp]
    \caption{Initial conditions.
    In the coplanar case, the longitude of the ascending node is not defined. 
    For the convenience of numerical integrations, we set the longitude of the ascending nodes as  $\Omega_\mathrm{in} = \Omega_\mathrm{out} = 0$.
    Outer argument of periapsis $\omega_\mathrm{out}$ and mean anomalies $M_\mathrm{in}, M_\mathrm{out}$ take the value $2\pi i/10$, where $i=0,1,...9$.}
    \centering
    \begin{tabular}{lcc} \hline
    Parameters & Symbol & Value \\ \hline
    mass of particles & $m_1, m_2, m_3$ & 0.5 $M_\odot$ \\
    inner semi-major axis & $a_\mathrm{in}$ & 1 au \\
    inner eccentricity & $e_\mathrm{in}$ & 0.5 \\
    outer eccentricity & $e_\mathrm{out}$ & 0.25 \\
    inner argument of periapsis & $\omega_\mathrm{in}$ & 0 \\
    outer argument of periapsis & $\omega_\mathrm{out}$ & $\mathcal{U}[0, 2\pi)$ \\
    inner mean anomaly & $M_\mathrm{in}$ & $\mathcal{U}[0, 2\pi)$ \\
    outer mean anomaly & $M_\mathrm{out}$ & $\mathcal{U}[0, 2\pi)$ \\ \hline
    \end{tabular}
    \label{tab:elements}
\end{table}

We set $a_\mathrm{out}$ so that $Q$ at the initial condition of each run is in the mixed-region.
By applying our initial parameters in Table \ref{tab:elements} to equation (\ref{eq:MA01}) , we can define the stability threshold as $Q \simeq 3.81$ for prograde orbits and $Q \simeq 2.67$ for retrograde orbits.
These two $Q$ values represent the stability limits for the coplanar cases determined from MA01 criterion.
So we use $Q=3.80, 3.77, 3.85, 3.73$ for prograde orbits, and $Q=2.60, 2.58, 2.55, 2.52$ for retrograde orbits.
These values of $Q$ are also arbitrary, but they lie in a region where stable and unstable systems coexist, making them appropriate for the purpose of this study.
Once $Q$ is determined, the outer semi-major axis ($a_\mathrm{out}$) can be calculated from equation (\ref{eq:q}), and all the orbital elements can be obtained.
When the $Q$ is determined, then $a_\mathrm{out}$ can be obtained using equation (\ref{eq:q}).

Our initial conditions consist of two types of orbital inclination: prograde and retrograde.
For each case, there are four different values of $Q$, along with three parameters that are uniformly distributed. 
Therefore, there are $4 \times 10 \times 10 \times 10 = 4000$ initial conditions for prograde orbits, and 4000 initial conditions for retrograde orbits as well.

\subsection{Numerical Method}\label{sec:211}
We integrate coplanar equal-mass triples for a maximum time of $10^9P_\mathrm{in}$, where $P_\mathrm{in}$ is the initial inner orbital period. 
We continue the integrations until the system disintegrates or the time reaches to $10^9P_\mathrm{in}$.
When the binding energy of either the inner or outer binary becomes negative, we regard the system as disintegrate.
The snapshots are recorded at intervals of $0.1P_\mathrm{in}$ for the first $10^3P_\mathrm{in}$. 
This interval is determined by balancing computational resources with the granularity of the data. 
By storing snapshots at such fine intervals, we can improve the analysis accuracy in the subsequent Fast Fourier Transform (FFT).

We used Algorithmic regularization (AR) for our integration, which is also called as Time-Transformed Symplectic Integrator (TSI) or LogH method \citep{PT99,MT99}.
AR is described in the extended phase space. 
Using AR, time is treated as one of the variables to be integrated in an extended phase space composed of time, positions, and velocities.
When applied to time integration of a two-body problem with the leap-frog method, AR gives the exact trajectory, the conservation of energy, and angular momentum of the system.
For these reasons, AR is well-suited for long-term integration of few-body systems.
We combine AR with a 6th-order symplectic formula by \citet{Yoshida90} to improve its accuracy.
As stated in \citet{sdar}, the desirable properties of AR are preserved even when combined with the Yoshida 6th-order symplectic method.
In this study, we modified the sample code included in the \textsc{sdar} library \footnote{ https://github.com/lwang-astro/SDAR} \citep{sdar} to create a numerical code for the AR+Yoshida6th method. 
\textsc{sdar} is an open-source library for integrating few-body systems, and it is available for anyone to use.
In addition to the AR+Yoshida6th method, we performed similar simulations with \textsc{tsunami} \citep{Trani+23} and confirmed that our results are independent of the integrator used.

\section{Results}
\subsection{The overall results of all integrations: the distribution of the survival time}

\begin{figure*}[htbp]
    \begin{center}
    \begin{minipage}{0.48\textwidth}
    \includegraphics[width=0.9\linewidth]{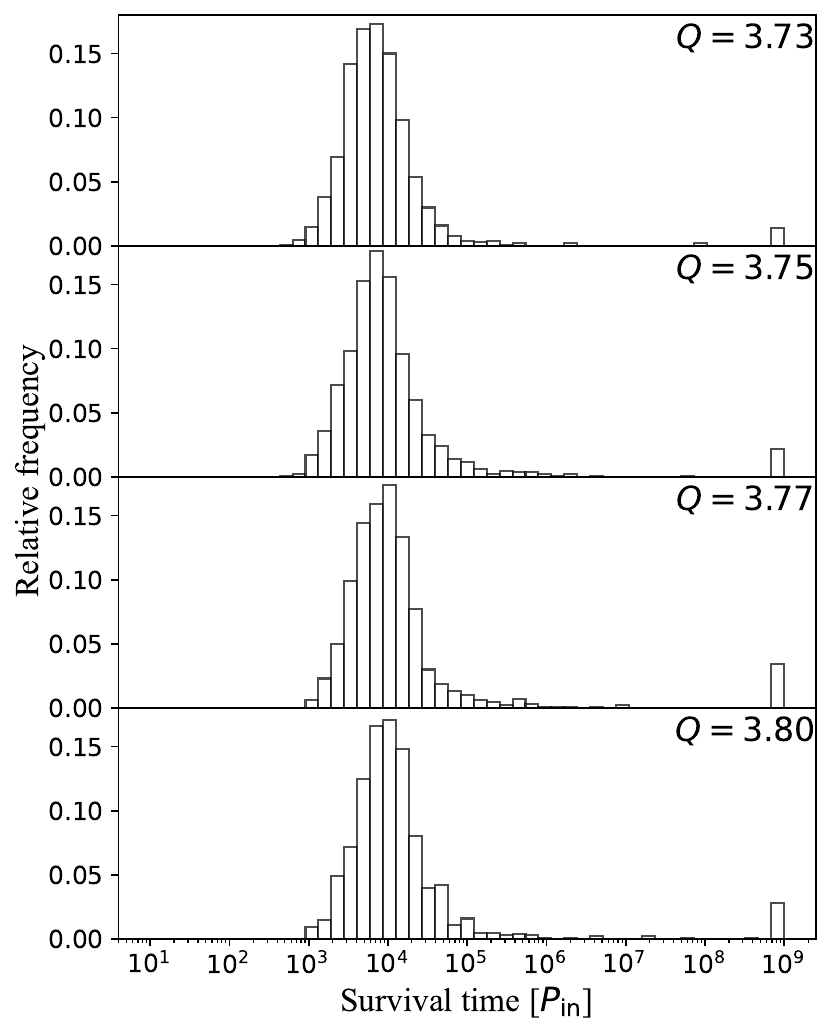}
    \end{minipage}
    \begin{minipage}{0.48\textwidth}
    \includegraphics[width=0.9\linewidth]{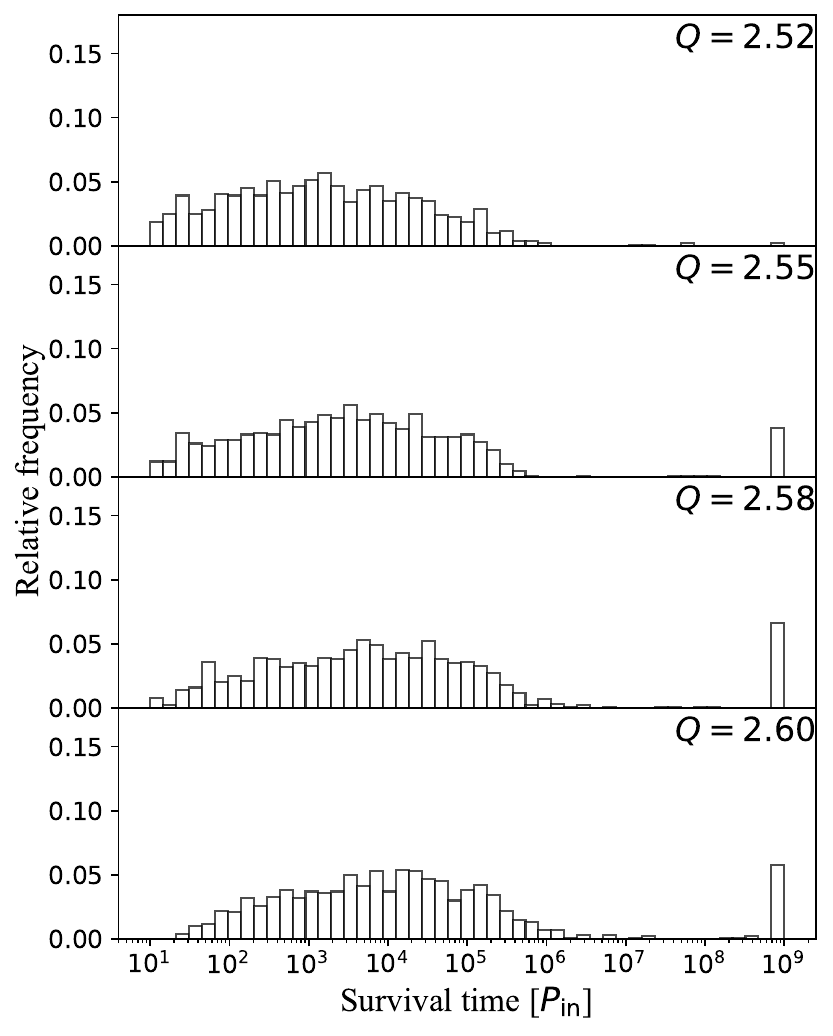}
    \end{minipage}
    \end{center}
    \caption{The distribution of survival time distribution for each $Q$.
    The left panel is prograde cases and the right panel is retrograde cases.
    For each cases, we distribute 50 bins between $10^1P_\mathrm{in}$ to $10^9P_\mathrm{in}$ on a logarithmic scale, and we plot the relative frequency of each bin. \\
    {ALT text: Histogram plots for each of the eight Q values.
    }}
    \label{fig:suvtime_distribution}
\end{figure*}

Figure \ref{fig:suvtime_distribution} shows the distribution of the survival time.
In the prograde case, the majority of systems break up in the period between $10^3P_\mathrm{in}$ and $10^6P_\mathrm{in}$. 
The number of systems with lifetimes between $10^7P_\mathrm{in}$ and $10^9P_\mathrm{in}$ is extremely small, and those surviving beyond $10^9P_\mathrm{in}$ account for approximately 1\% to 4\% of the total.
In the retrograde case, the distribution is broader than that of the prograde case, with most systems breaking up in the period between $10^2P_\mathrm{in}$ and $10^6P_\mathrm{in}$.  
Systems surviving beyond $10^9 P_\mathrm{in}$ account for approximately 5\% of the total, except for the case of $Q=2.52$.

As shown in figure \ref{fig:suvtime_distribution}, even systems with the same $Q$ exhibit variations in survival time of several orders of magnitude.
Our initial conditions vary only the argument of periapsis ($\omega$) and initial phases for a given $Q$, so the combination of $\omega$ and phases produces the distribution observed in figure \ref{fig:suvtime_distribution}.
The current stability conditions incorporate dependencies on the mass, semi-major axis, eccentricity, and orbital inclination, but the dependence on $\omega$ and phases has not yet been thoroughly investigated.
Therefore, the distribution of stability and survival time observed in figure \ref{fig:suvtime_distribution} cannot be captured at all by the current stability conditions.

\subsection{Periodicity of orbital evolution\label{sec:orbital_evolution}}
\begin{figure*}[htbp]
    \begin{center}
    \begin{minipage}{0.48\textwidth}
    \includegraphics[width=\linewidth]{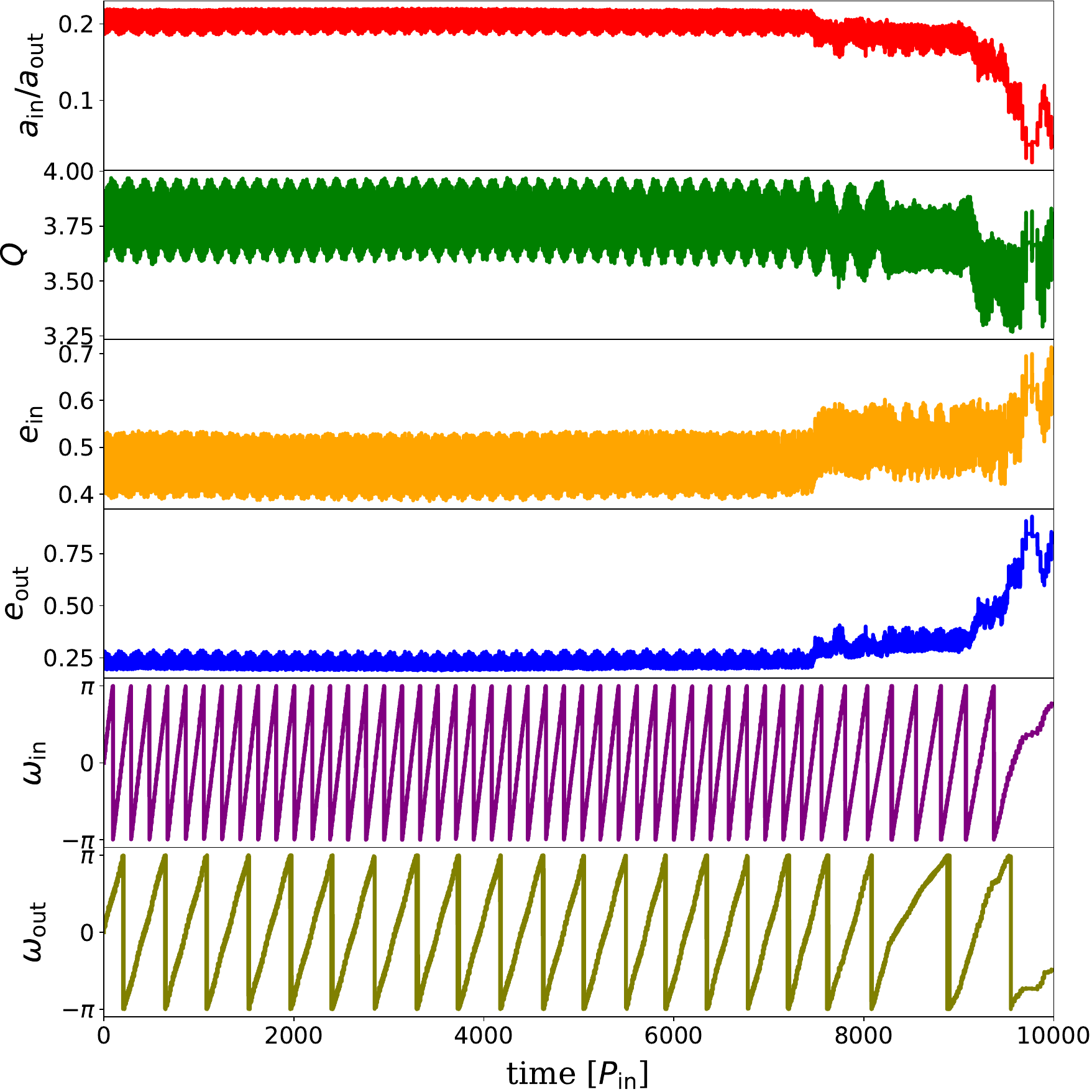}
    \end{minipage}
    \begin{minipage}{0.48\textwidth}
    \includegraphics[width=\linewidth]{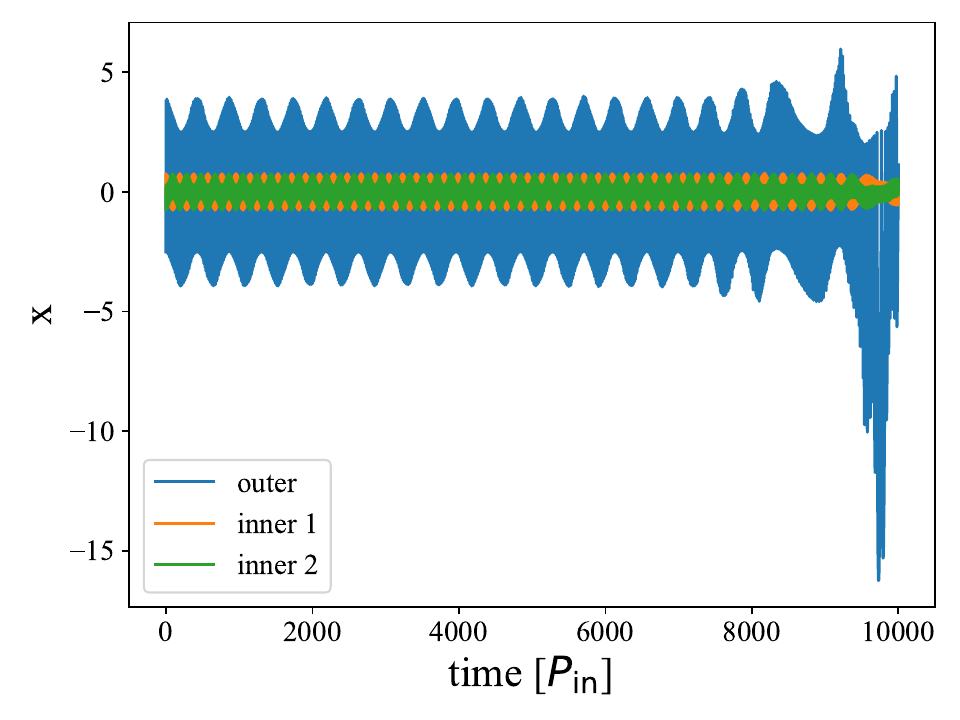}
    \end{minipage}
    \end{center}
    \caption{The orbital evolution of an unstable prograde orbit with $Q = 3.75$ and the time evolution of the $x$-coordinates of the three bodies. \\
    {ALT text: Time evolution of the orbital elements and the x-coordinates of the three bodies.}}
    \label{fig:app_element_and_x_evo}
\end{figure*}

In this section, we compare the orbital evolution of stable and unstable systems shown in figure \ref{fig:suvtime_distribution}. 
First, we explain the mechanism of system destabilization.
Figure \ref{fig:app_element_and_x_evo} shows the time evolution of orbital elements and the time evolution of the $x$-coordinates of each of the three bodies.
Note that in our calculations the orbit is in the $x$-$y$ plane.
In this case, the three-body system breaks down at approximately $10500P_\mathrm{in}$, but since this study only collects detailed output up to $10000P_\mathrm{in}$, the data is shown up to $10000P_\mathrm{in}$, which is sufficient for our explanation.
The orbital evolution shows that, initially, the system evolves in a relatively periodic manner up to approximately $7000P_\mathrm{in}$. 
However, around $8000P_\mathrm{in}$, fluctuations in the orbital elements occur, which contribute to the destabilization of the system. 
Eventually, the outer orbit becomes highly eccentric.
Systems that reach this state eventually undergo disruption.

\begin{figure}
    \begin{center}
    \includegraphics[width=\linewidth]{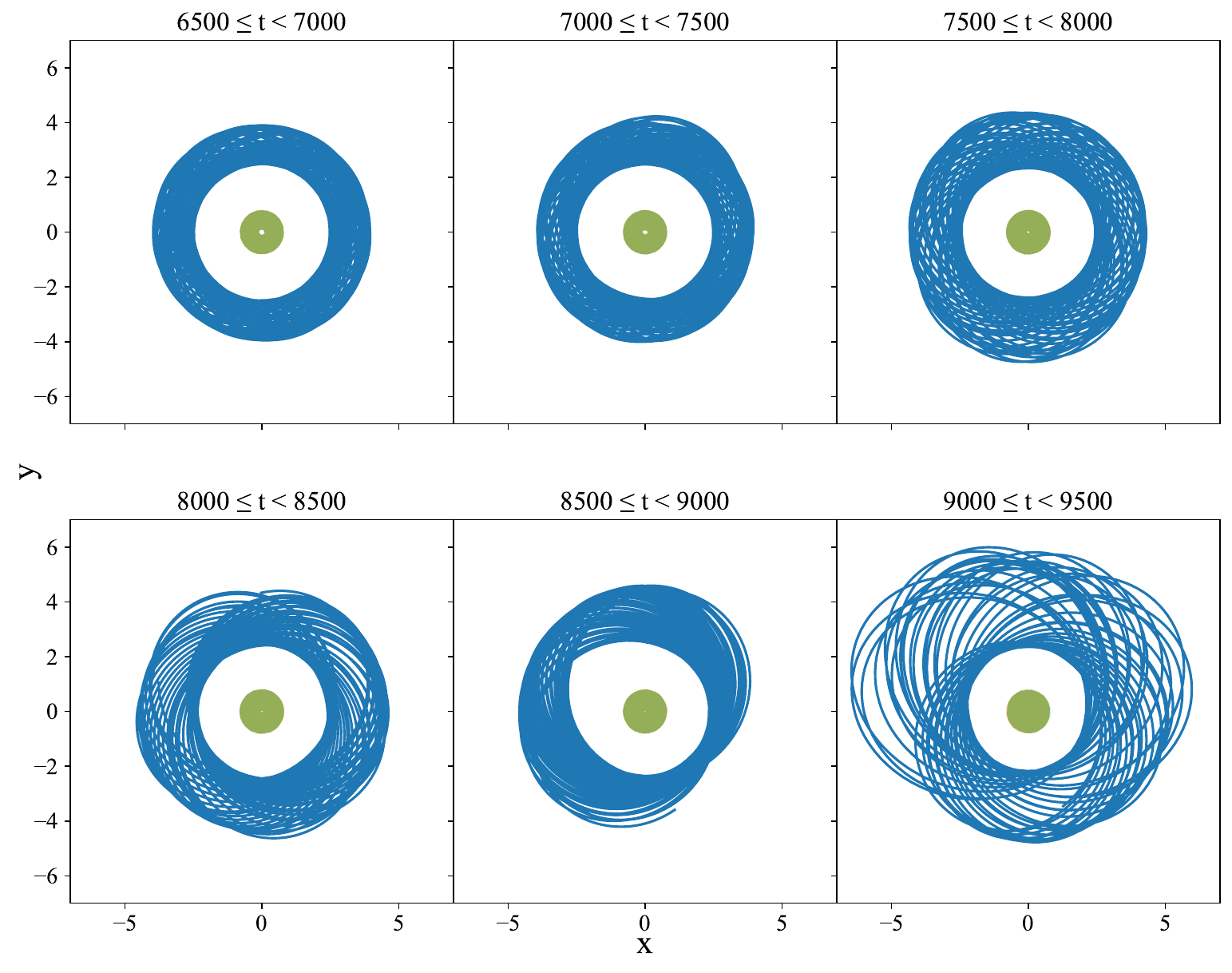}
    \end{center}
    \caption{The trajectories of the three bodies on the orbital plane from $t=6500P_\mathrm{in}$ to $t=9500P_\mathrm{in}$. \\
    {ALT text: Six snapshots depicting the trajectory of the orbit.}}
    \label{fig:xy}
\end{figure}

Figure \ref{fig:xy} shows the trajectories of the system from figure \ref{fig:app_element_and_x_evo} as viewed in the orbital plane ($x$-$y$ plane). 
The initial orbital variations occur between $7000P_\mathrm{in}$ and $8000P_\mathrm{in}$ (upper center and right in figure \ref{fig:xy}), during which it is evident that the shape of the outer orbit undergoes significant changes before and after this time.

Based on the above considerations, it is anticipated that stable orbits will not undergo significant variations over time. 
This raises the question: what distinguishes stable orbits from unstable ones?
In the following, we conduct a comparative analysis between stable and unstable orbits in order to elucidate their differences.

\begin{figure}[htbp]
    \begin{center}
    \includegraphics[width=\columnwidth]{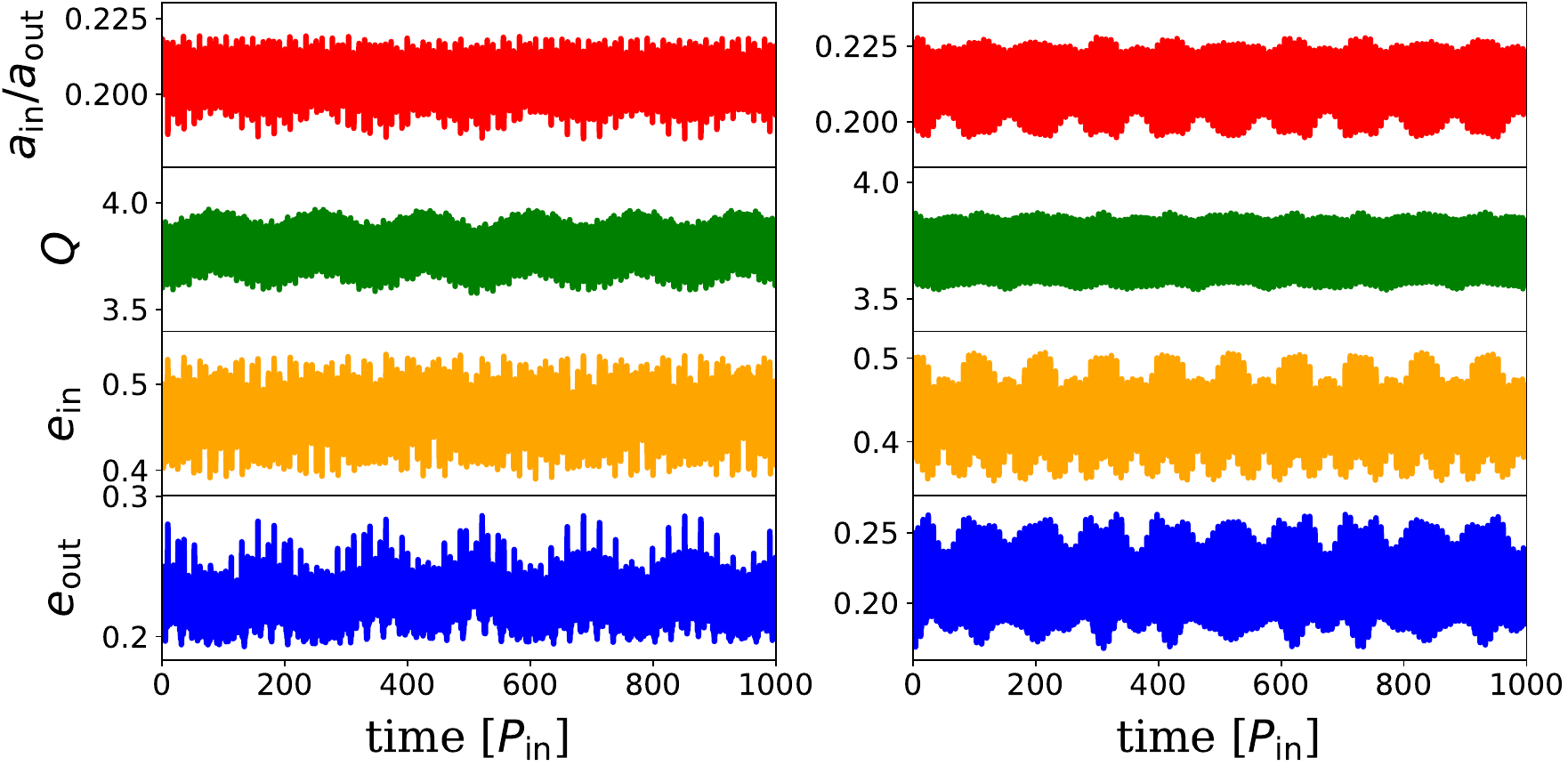}
    \end{center}
    \caption{Examples of orbital elements evolution: the left is for an unstable system, and right is for a stable system.
    These are for the cases with $Q=3.75$. \\
    {ALT text: A figure comparing four orbital elements between stable and unstable orbits.
    Four orbital elements are the semi-major axis ratio, Q, inner eccentricity, and outer eccentricity.}
    }
    \label{fig:element_comparison}
\end{figure}

Figure \ref{fig:element_comparison} shows the orbital evolution over the first $1000 P_{\mathrm{in}}$ of the systems shown in figure \ref{fig:app_element_and_x_evo} and that of a stable system having the same value of $Q=3.75$.
The unstable system (the left-hand side panel) exhibits significant irregular variations in all orbital elements while irregularities seem smaller for the stable system(right-hand side panel).
This difference might imply that there exists a meaningful relationship between the degree of the periodicity of the orbital elements and the long-term dynamical stability of the system.

In figure \ref{fig:element_comparison}, we just give one particularly illustrative example.
We need some measure for this degree of periodicity.
Therefore, we employ the Fast Fourier Transform (FFT) to quantitatively investigate the relationship between periodicity and dynamical stability.
In this paper, as the first trial, we focus on the evolution of the ratio of semi-major axes of the inner and outer orbits ($a_\mathrm{in}/a_\mathrm{out}$).

Here, we present our procedure to apply FFT to a measured quantity.
Before performing the FFT, the mean value of the 8192 data points is calculated and subtracted from each data point. 
This step is done to minimize the DC component as much as possible.
After this process, a Hanning window function is applied to the data, and then the FFT is performed. 
The resulting amplitudes are normalized by dividing them by half the total number of data points, i.e., 4096.
In this study, we output the orbital elements with a $0.1P_\mathrm{in}$ interval for the first $10^3P_\mathrm{in}$.
Since the FFT requires the number of data points to be a power of 2, we extracted the first 8192 data points, corresponding to $819.2P_\mathrm{in} \sim 10^3P_\mathrm{in}$, and performed the FFT.
The sampling frequency of the FFT is 10, and the Nyquist frequency is 5.
We discarded systems that survive less than $819.2P_\mathrm{in}$, since we cannot do FFT analysis on such systems for sufficient time.
There are 5 prograde systems and 1287 retrograde systems whose survival times did not reach  $819.2P_\mathrm{in}$, so we present the results for 3995 prograde cases and 2713 retrograde cases. 

\begin{figure*}[htbp]
    \begin{center}
    \includegraphics[width=0.93\textwidth]{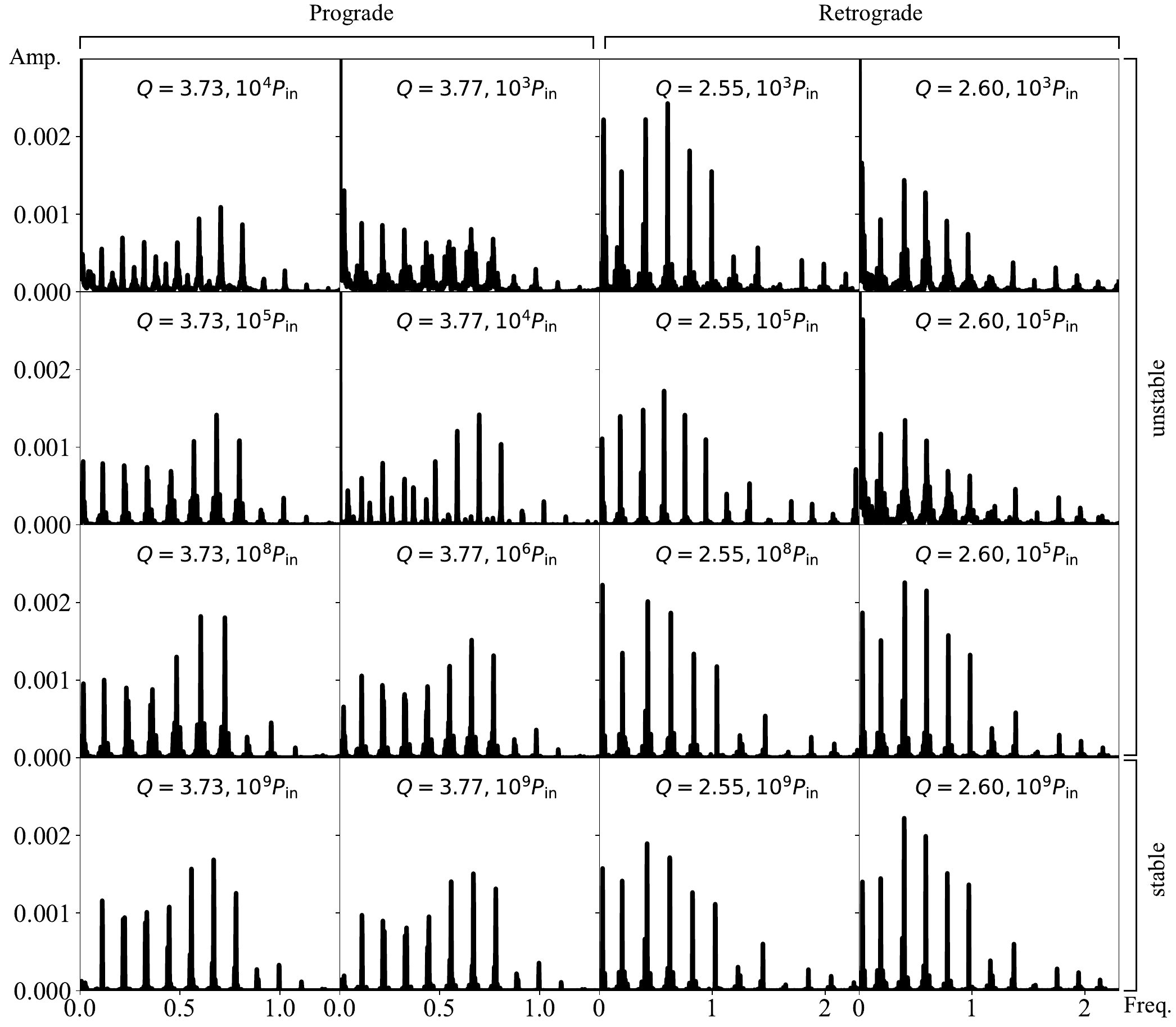}
    \end{center}
    \caption{The frequency distribution of orbital evolution ($a_\mathrm{in}/a_\mathrm{out}$). 
    The horizontal and vertical axes correspond to the frequency and the amplitude, respectively.
    From left to right, each column shows 4 cases for prograde ($Q = 3.73$), 4 prograde cases ($Q=3.77$), 4 retrograde cases ($Q=2.55$), and 4 retrograde cases ($Q=2.60$).
    The survival time of the corresponding system is shown in the upper right of each panel. \\
    {ALT text: Sixteen panels showing the results of the FFT of the orbits.
    Stable orbits exhibit distinct, evenly spaced peaks.
    In unstable cases, strong continuous components are present, and peaks are not evenly spaced.}
    }
    \label{fig:fft_many}
\end{figure*}

Figure \ref{fig:fft_many} shows 16 examples of our FFT results.
The panels are arranged from top to bottom in the order of survival time for four different values of $Q$ (two prograde and two retrograde).
If the orbital evolution is perfectly periodic, only the fundamental frequency ($f_0$) and its integer multiples ($2f_0, 3f_0, \dots$) would appear in the frequency domain with a small effect of the window function which appears as small broadening of each peak and low-frequency terms.
The orbital evolution of the system with evenly spaced peaks and fewer continuous components, which are side-lobes of the peaks, therefore, is more stable.
When we compare systems with different lifetimes (for instance, comparing the top and bottom columns), it is evident that the stable systems exhibit more pronounced peaks with smaller side-lobes and non-periodic ``noises''.
This trend indicates that stable systems are more periodic.
This statement might sound almost like a tautology, since by definition periodic systems are stable.
However, as far as we know this is the first time that the numerical determination of periodicity is applied to the stability of hierarchical triples.
Additionally, a consistent feature across almost all cases is the presence of a noticeable peak near zero frequency.
This peak is not the fundamental frequency and its strength seem to be related to the early disintegration.

\begin{figure}[htbp]
    \begin{center}
    \includegraphics[width=\columnwidth]{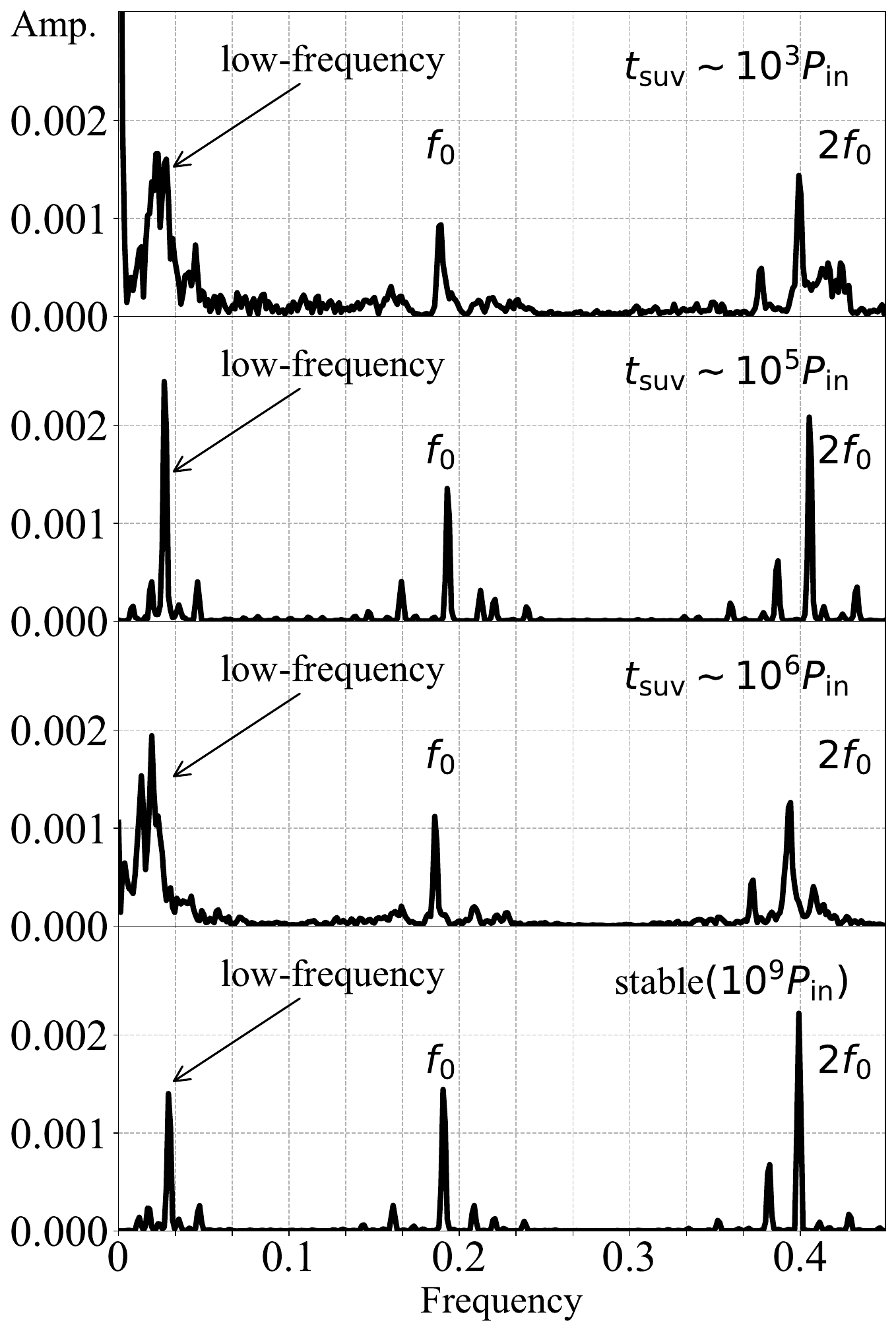}
    \end{center}
    \caption{4 examples of orbital evolution FFT for retrograde $Q=2.60$ cases.
    These figures correspond to the rightmost panel of figure 
    \ref{fig:fft_many}.
    The vertical axes represent the amplitude of FFT, while the horizontal axis represents frequency.
    The grid lines on the x-axis are evenly spaced at intervals of 1/30 units. \\
    {ALT text: Zoomed-in FFT spectra up to a frequency of 0.45 for four cases with Q equal 2.6.}}
    \label{fig:fft_several}
\end{figure}

Figure \ref{fig:fft_several} depicts examples of the FFT results ($Q=2.60$, the rightmost panel of figure \ref{fig:fft_many}), focusing on the frequency range up to 0.45.
The four cases in figure \ref{fig:fft_several} all correspond to retrograde cases, but a similar tendency is observed for prograde motion or cases with different $Q$ values.

First, we clarify the physical meaning of the peaks.
There are three peaks in figure \ref{fig:fft_several}, which correspond to two different types of variations.
The middle peak and the right peak originate from the same variation.
The peaks around 0.2 represent the fundamental frequencies ($f_0$), and the peaks observed around 0.4 correspond to the second harmonic of the fundamental frequency ($2f_0$).
The other peaks observed in figure \ref{fig:fft_many} are also components that are integer multiples of the fundamental frequency.
For $Q = 2.6$ , the period of the outer orbit is approximately five times that of the inner orbit.
Given that the frequency unit in the FFT is  $1/P_\mathrm{in}$ , the value $1/5 = 0.2$ indicates that these peaks are associated with variations caused by the periapsis passage of the outer orbit.
In contrast, the leftmost low-frequency peak originates from a different dynamical mechanism.
This peak represents the long-term variations caused by the continuous interaction between the inner and outer orbits such as mean-motion or secular resonances.
In this paper, the continuous components and peaks in the low-frequency region, as depicted in figure \ref{fig:fft_several}, are collectively referred to as {\it low-frequency components}.

Next, we compare the unstable ones (the top three) with the stable ones (the bottom).
As mentioned above, the system survives longer when peaks at $f_0$ and  $2f_0$ are clearly visible (the bottom of Fig \ref{fig:fft_several}).
On the other hand, the top three panels in figure \ref{fig:fft_several} show larger continuous components compared to the bottom panel (the most stable one). 
Such continuous components are characteristic of unstable systems.
Additionally, the first and third panels from the top show a DC component at zero frequency.
If the orbital evolution is not periodic, the center of oscillation deviates from the mean value of the data, resulting in the appearance of a DC component and peaks around zero frequency. 

The difference between the second panel from the top and the bottom panel in figure \ref{fig:fft_several} is also rather clear. 
It can be seen that the second panel from the top has clearly defined peaks, but the low-frequency component is larger than other peaks.
Though this feature is common in most cases, there are exceptions such as the second-to-top panel of figure \ref{fig:fft_many} with $Q = 2.55$.
Therefore, it is not possible to make a definitive conclusion.
Nevertheless, it is plausible that low-frequency components act as the dominant factor, which governs the stability.

To summarize, the FFT analysis of the orbital element evolution of a stable system reveals distinct fundamental frequency components and their integer multiples, with small continuous components. 
Furthermore, the peak near zero frequency is relatively small. 
It is rather surprising that such trends can be recognized by observing the orbital evolution for only $10^3 P_{\mathrm{in}}$. 

\begin{figure}[htbp]
    \begin{center}
    \begin{minipage}{0.48\textwidth}
    \includegraphics[width=\linewidth]{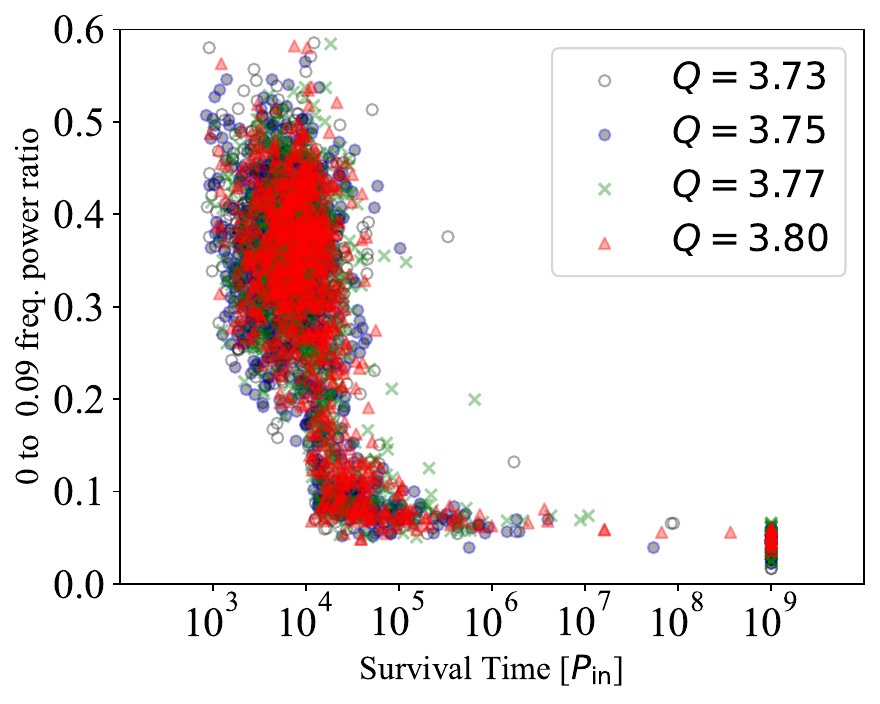}
    \end{minipage} \\
    \begin{minipage}{0.48\textwidth}
    \includegraphics[width=\linewidth]{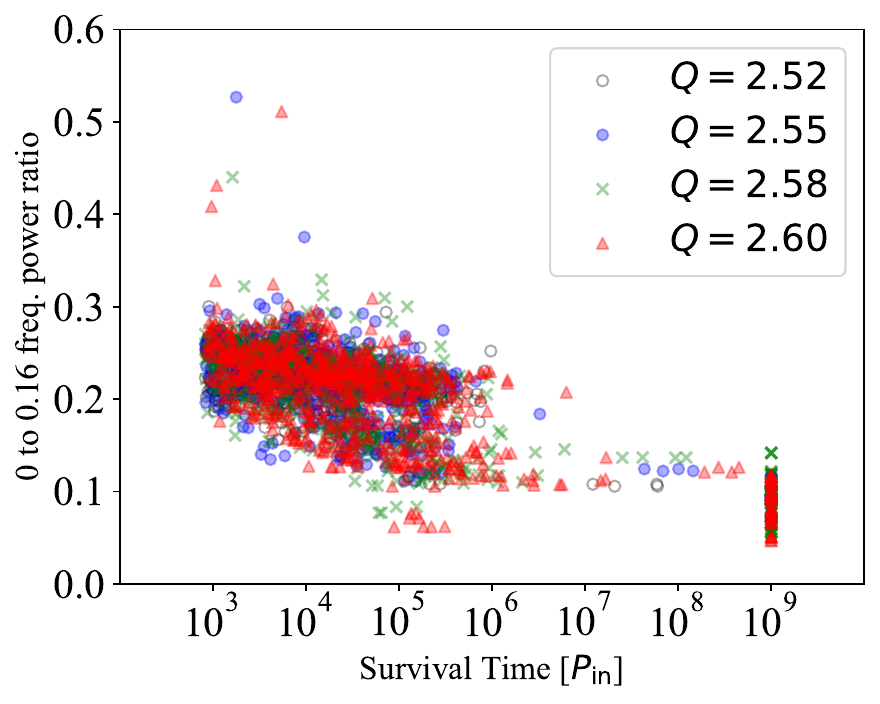}
    \end{minipage}
    \end{center}
    \caption{Relation between survival time and low-frequency components power ratio.
    The upper panel is for prograde orbits and the lower panel is for retrograde orbits. \\
    {ALT text: Two scatter plots showing the survival time versus the fraction of low-frequency components.}}
    \label{fig:time_to_power_ratio}
\end{figure}

\subsection{Probability analysis of stability prediction using FFT result \label{sec:quanty_and_prefict}}
\subsubsection{Quantification of Periodicity}
In this section, we quantitatively evaluate the periodicity of the orbital evolution and show its relationship with the stability.
In the previous section, we found that stable triple systems tend to have strong peaks at the fundamental frequency and its integer multiples, with minimal continuous components.
Systems that disintegrate tend to have continuous components and/or a large peak between zero and the fundamental frequencies.
These components seem to be related to short lifetimes.

We focused on this {\it low-frequency component} in order to quantify the non-periodicity of the orbital evolutions.
Our FFT results show that the fundamental frequency appears around 0.10 for prograde systems and around 0.19 for retrograde systems.
Taking this into account, we calculated the sum of the power of frequencies lower than this fundamental frequency and expressed it as a ratio to the total power of all peaks.
For the prograde case, components with frequencies below 0.09 were regarded as low-frequency components.
For the retrograde case, components with frequencies below 0.16 were regarded as low-frequency components.

The power ratio of this low-frequency component is plotted against the triple's survival time in figure \ref{fig:time_to_power_ratio}.
There is a clear correlation between the power of the low-frequency component and the lifetime of the system.
Note that this low-frequency component is not a single factor that determines the lifetime.
We can see that there are some systems with small low-frequency power which still exhibit short lifetimes.
Even so, it is clear that we can use this measure as one way to predict the lifetime.

\subsubsection{Predictability}
In the previous subsection, we proposed a method to predict stability using FFT results.
Figure \ref{fig:time_to_power_ratio} illustrates the relationship between the system’s survival time, which corresponds to its stability, and the magnitude of its low-frequency component. in the time variation of the orbital element.
We classified systems with a specific power ratio.
The systems with a power ratio smaller than the threshold are regarded as stable.
For example, in the case of retrograde orbits shown in figure \ref{fig:time_to_power_ratio}, drawing a horizontal line at a power ratio of 0.15 allows for the detection of all systems that survive above 
$10^9P_\mathrm{in}$.
When the power ratio is divided at a certain value, four patterns can be observed: 
(i) a stable system is classified as stable (true positive; TP), 
(ii) an unstable system is classified as stable (false positive; FP), 
(iii) an unstable system is classified as unstable (true negative; TN), 
and (iv) a stable system is classified as unstable (false negative; FN).
We verify the validity of our results and the accuracy of the predictions using these four classifications.

\begin{figure}[htbp]
    \begin{center}
    \begin{minipage}{0.48\textwidth}
    \includegraphics[width=\columnwidth]{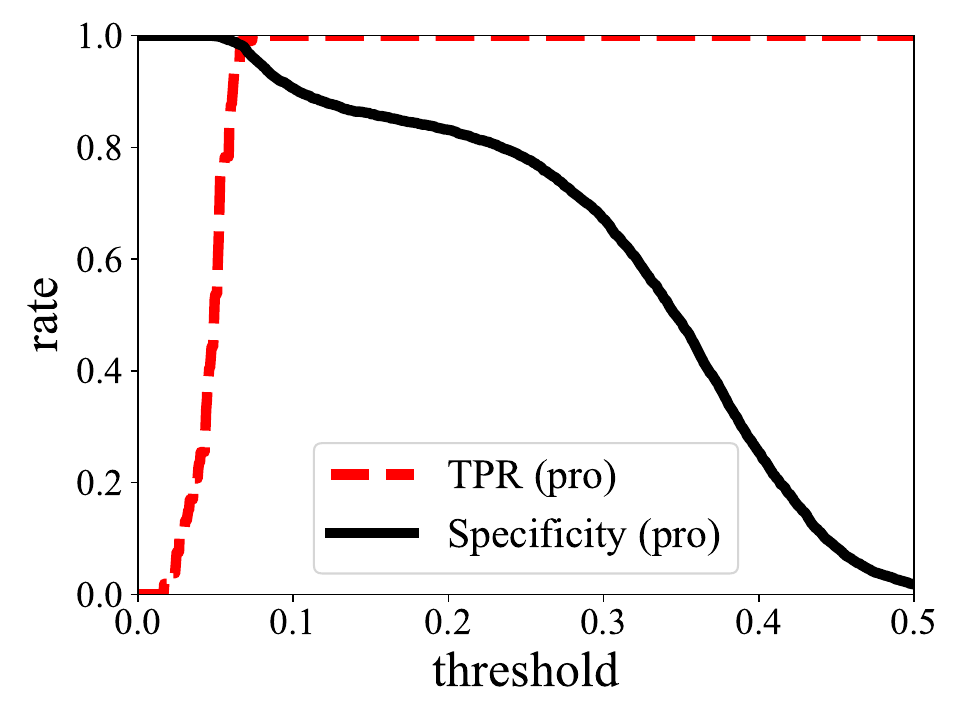}
    \end{minipage} \\
    \begin{minipage}{0.48\textwidth}
    \includegraphics[width=\columnwidth]{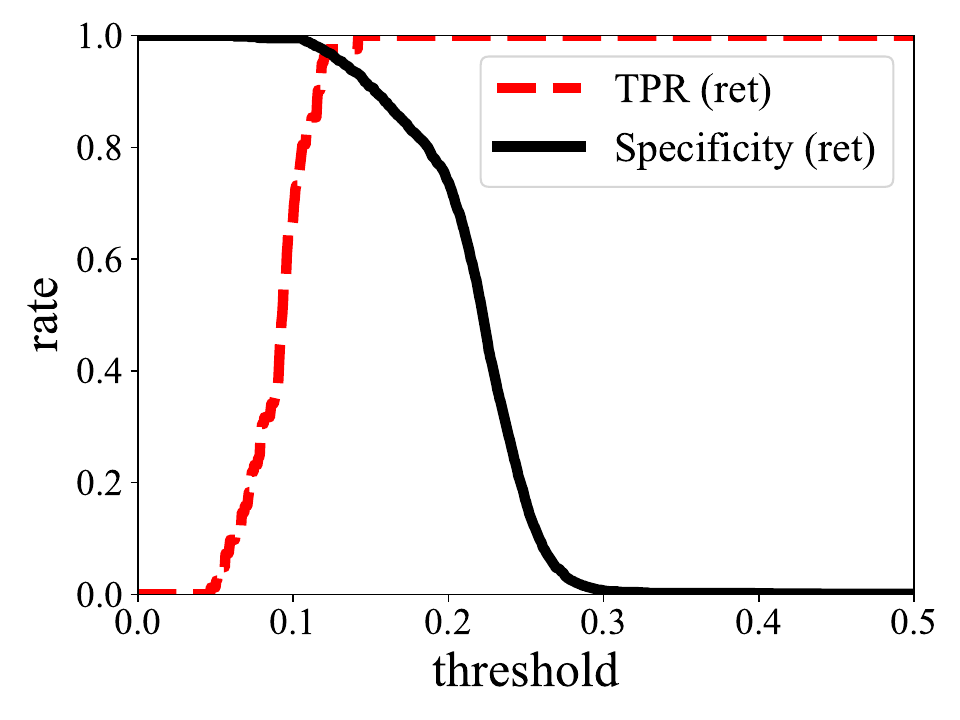}
    \end{minipage}
    \end{center}
    \caption{The changes in TPR and Speficifity (1-FPR) with respect to the threshold. 
    The upper panel shows the case for prograde orbits, while the lower panel shows the case for retrograde orbits. \\
    {ALT text: Two line plots of the decision threshold versus true positive rate and specificity.}} 
    \label{fig:tpr_fpr}
\end{figure}

Before proceeding to the detailed evaluation of accuracy, we introduce the following metrics:
\begin{enumerate}
    \item Accuracy: the ratio of the number of correct predictions to the total number of systems: (TP + TN)/(TP + TN + FP + FN).
    \item Precision: the ratio of the number of actual stable systems to that of systems regarded as stable: TP / (TP + FP).
    \item Recall or True Positive Rate (TPR): the ratio of the number of systems correctly identified as stable to that of real stable systems: TP / (TP + FN).
    \item False Positive Rate (FPR): the ratio of the number of systems that are incorrectly classified as stable to the total number of real unstable systems: FP / (TN + FP).
    \item Specificity: the ratio of the number of systems correctly classified as unstable to that of unstable systems: 1-FPR.
    \item F-measure: the harmonic mean of Precision and Accuracy.
\end{enumerate}
These metrics vary depending on the threshold.
Figure \ref{fig:tpr_fpr} shows the TPR and Specificity as functions of the threshold.
Ideally, the TPR and Specificity should be as close to 1 as possible.
In practice, the choice of a threshold must be determined by the user based on what type of problem the stability criterion is intended to address.

\begin{figure*}[h!]
    \begin{center}
    \begin{minipage}{0.48\textwidth}
    \includegraphics[width=\columnwidth]{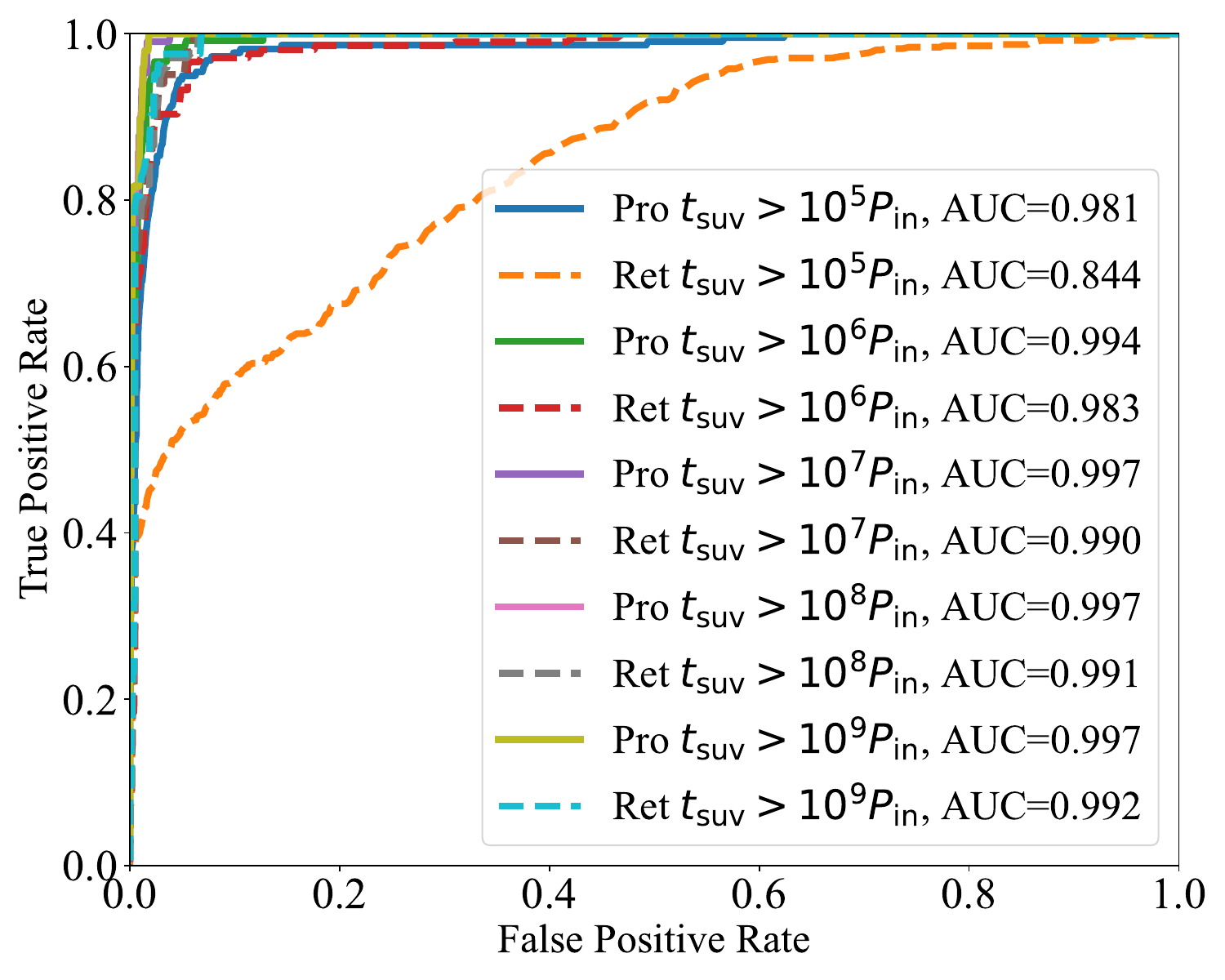}
    \end{minipage}
    \begin{minipage}{0.48\textwidth}
    \includegraphics[width=\columnwidth]{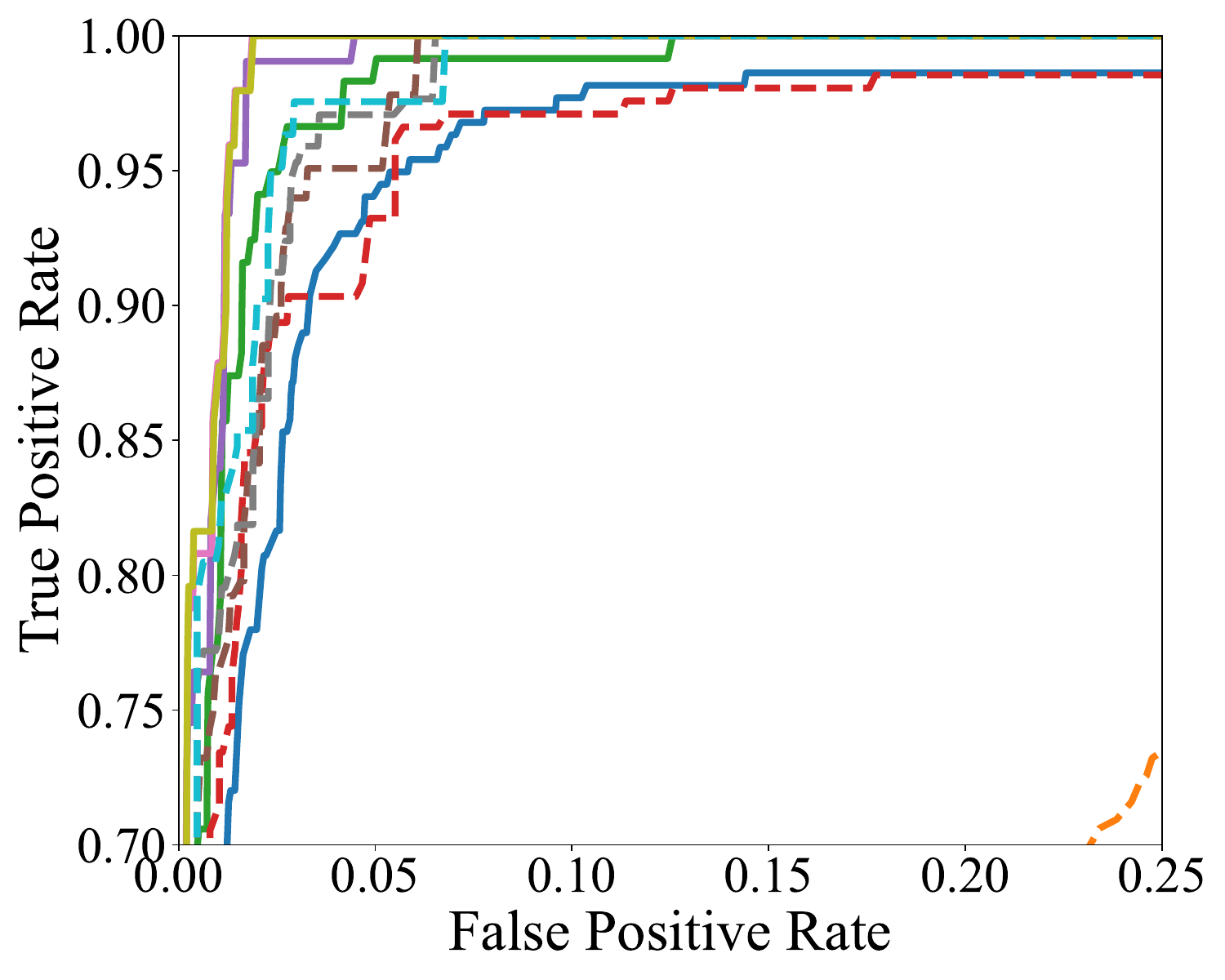}
    \end{minipage}
    \end{center}
    \caption{ROC curves of prograde and retrograde cases.
    Each lines correspond to different survival time ($t_\mathrm{suv}$) and inclination (prograde and retrograde).
    The left panel is a overall look, and the right panel is a closer look at the upper-left corner of the left panel. \\
    {ALT text: ROC curve and its zoomed-in view.}} 
    \label{fig:roc}
\end{figure*}

First, we assess the performance of our method of identification of stability of hierarchical triples using power spectrum with receiver operating characteristic (ROC) curves.
The ROC curve visualizes the TPR values as the threshold is progressively increased, thereby adjusting the FPR from 0 to 1.
Figure \ref{fig:roc} shows ROC curves for our results.
The horizontal axis represents the False Positive Rate (FPR), and the vertical axis represents the True Positive Rate (TPR).
The definition of “stability” is based on systems that survive for a certain minimum period of time ($t_\mathrm{stable}$). 
We evaluated for the cases of $t_\mathrm{stable} = 10^5P_\mathrm{in}$, $10^6P_\mathrm{in}$,  $10^7P_\mathrm{in}$, $10^8P_\mathrm{in}$, and $10^9P_\mathrm{in}$.
The numbers in the legend represent the value of area under the curve (AUC).
The AUC is over 98\% for distinguishing systems stable for more than $10^6P_\mathrm{in}$, and over 99\% for identifying systems stable for more than $10^7P_\mathrm{in}$.
As shown in the survival time distribution in figure \ref{fig:suvtime_distribution}, the majority of systems are destroyed relatively early, while systems with survival times exceeding $10^7P_\mathrm{in}$ deviate from the majority.
These AUC values indicate that the orbital evolution of systems with a lifetime longer than $10^7P_\mathrm{in}$ looks rather periodic and thus the low-frequency component is small.

Though AUC gives one measure of the predictive power, its meaning is a bit difficult to interpret since it does not directly give, for example, the probability that that the prediction is correct.
So we present the values of metrics which directly gives such probabilities, for one particular value of the threshold.
We chose the value of the threshold so that the F-measure is the maximum.
Since the Precision and Recall are in a trade-off relationship, a maximum F-measure indicates a well-balanced criterion between them.
Table \ref{tab:metrics} shows the values of performance metrics we mentioned above.
The parameter $t_\mathrm{stable}$ is the lifetime of the system above which we regard it as stable.
The maximum F-measure is approximately 0.8.
We can see that the Accuracy is very high, 0.98--0.99.
This means that only 1--2\% of the predictions are wrong, when we consider predictions for both the stable and unstable cases.
If we look at only the prediction that the system is stable, the precision is around 0.9 and recall rate is around 0.7.
These values are much lower than the Accuracy simply because 95--97\% of the systems are unstable.

Table \ref{tab:metrics_unstab} shows the value of metrics for the prediction of unstable systems.
By definition, the accuracy is identical between Table \ref{tab:metrics} and Table \ref{tab:metrics_unstab}.
When instability is treated as the positive class, the maximum F-measure becomes higher—around 0.99—compared to the case where stability is treated as positive. 
This is because the majority of the systems we prepared become unstable within $10^6 P_\mathrm{in}$, and are therefore labeled as unstable.
As will be shown in the discussion section, the predictive performance of our method is higher than that of previously proposed approaches.

\begin{table*}[h!]
\caption{Metrics for the prediction of stable systems.
The definitions of each parameter are provided in the main text.}
\label{tab:metrics}
\centering
\begin{tabular}{llccccccc}
\hline
& $t_{\mathrm{stable}}$ & threshold & max F-measure & Accuracy & Precision & Recall (TPR) & FPR & Specificity \\
\hline
Prograde & $10^7 P_\mathrm{in}$ & 0.054 & 0.827 & 0.991 & 0.900 & 0.764 & 0.002 & 0.998 \\
& $10^8 P_\mathrm{in}$ & 0.054 & 0.847 & 0.993 & 0.889 & 0.808 & 0.003 & 0.997 \\
& $10^9 P_\mathrm{in}$ & 0.054 & 0.851 & 0.993 & 0.889 & 0.816 & 0.0/03 & 0.997 \\
\hline
Retrograde & $10^7 P_\mathrm{in}$ & 0.106 & 0.810 & 0.977 & 0.905 & 0.732 & 0.006 & 0.994 \\
& $10^8 P_\mathrm{in}$ & 0.105 & 0.823 & 0.980 & 0.914 & 0.749 & 0.005 & 0.995 \\
& $10^9 P_\mathrm{in}$ & 0.105 & 0.842 & 0.982 & 0.914 & 0.780 & 0.005 & 0.995 \\
\hline
\end{tabular}
\end{table*}

\begin{table*}[h!]
\caption{Metrics for the prediction of unstable systems.}
\label{tab:metrics_unstab}
\centering
\begin{tabular}{llccccccc}
\hline
& $t_{\mathrm{stable}}$ & threshold & max F-measure & Accuracy & Precision & Recall (TPR) & FPR & Specificity \\
\hline
Prograde & $10^7 P_\mathrm{in}$ & 0.054 & 0.996 & 0.991 & 0.994 & 0.998 & 0.236 & 0.764 \\
& $10^8 P_\mathrm{in}$ & 0.054 & 0.996 & 0.993 & 0.995 & 0.997 & 0.192 & 0.808 \\
& $10^9 P_\mathrm{in}$ & 0.054 & 0.996 & 0.993 & 0.995 & 0.997 & 0.184 & 0.816 \\
\hline
Retrograde & $10^7 P_\mathrm{in}$ & 0.106 & 0.988 & 0.977 & 0.981 & 0.994 & 0.268 & 0.732 \\
& $10^8 P_\mathrm{in}$ & 0.105 & 0.989 & 0.980 & 0.983 & 0.995 & 0.251 & 0.749 \\
& $10^9 P_\mathrm{in}$ & 0.105 & 0.991 & 0.982 & 0.986 & 0.995 & 0.220 & 0.780 \\
\hline
\end{tabular}
\end{table*}

In $N$-body simulations, it is essential to identify stable systems as stable to reduce the computational cost of three-body calculations, without misidentifying unstable ones as stable.
For unstable systems, the Recall rate is related to the efficiency of $N$-body simulation, while Precision relates to the reliability. 
Therefore, clearly we need a prediction method which gives a high Recall rate while keeping Precision high close to unity for unstable systems, and thus F-measure for unstable system identification is a good measure.

\section{Discussion}
\subsection{Comparison with previous works}
Our findings demonstrate that using $Q$ to determine stability in the mixed region is inadequate, even though this approach is widely employed in many $N$-body simulation codes. We integrated approximately 8000 systems with initial values of $Q$ about 5\% smaller than the critical value from MA01, and found that around 5\% of these systems remained stable, defined as having a lifetime longer than $10^9 P_\mathrm{in}$.

First, we compare our method with that of V22, one of the most accurate approaches among those utilizing $Q$ to date. V22 proposed two types of stability criteria: one is an improved version of the MA01 criterion, and the other enhances accuracy by incorporating machine learning. Both criteria make stability predictions solely based on initial parameters, and their design does not fundamentally differ from conventional approaches. The stability classification results using these criteria are summarized in Table 4 of V22, where the original MA01 criterion achieves an accuracy of 90\%, the improved analytical criterion reaches 93\%, and the combination with machine learning achieves 95\%.

However, it is important to note that V22 evaluates stability over a broad parameter space. In particular, the ratio of the inner to outer semi-major axes spans $10^{-4} < a_\mathrm{in}/a_\mathrm{out} < 1$. This range is considerably broader than the mixed region and includes domains where stability or instability can be trivially determined. Therefore, the improvement in accuracy—from 90\% to 95\%—is attributed to better fitting of the overall shape of the mixed region, rather than enhanced classification within the region itself. We interpret the remaining 5\% misclassification rate as arising from ambiguity within the mixed region, where stability is intrinsically difficult to determine using only $Q$.

In contrast, our method achieves an accuracy of approximately 98–99\% specifically within the mixed region—an area where conventional $Q$-based criteria fail to make reliable predictions. By separating the parameter space into domains where traditional $Q$-based criteria can easily classify stability and those within the mixed region, and applying our method to the latter, we demonstrate a substantial improvement in overall accuracy compared to existing approaches.

Second, we compare our method with LT22, which presents a stability criterion based on orbital evolution rather than solely on initial conditions. LT22 uses a machine learning approach to predict the lifetime from variations in orbital elements, achieving an AUC score of around 95\%. They evaluate stability by sampling the early phase of orbital evolution up to $5 \times 10^3 P_\mathrm{in}$ at intervals of $5\pi P_\mathrm{in}$. In contrast, we sample the evolution at much finer intervals of $0.1 P_\mathrm{in}$—approximately 100 times denser—while requiring only a short integration time of about $10^3 P_\mathrm{in}$ to make a prediction. As a result, the computational cost of our stability classification is significantly reduced.

Our method achieves an AUC score exceeding 99\%, which is notably higher than the approximately 95\% reported by LT22. A comparison between figure 5 in LT22 and our figure \ref{fig:roc} highlights a distinct difference in behavior near the top-left corner of the ROC curves: our method maintains a TPR of 100\% while keeping the FPR below 10\% (figure \ref{fig:roc}).

In summary, even with improved fitting of the stability criterion, the performance of the conventional $Q$-based approaches is not very high.
In the mixed-region, it is more effective to incorporate orbital evolution into the classification process, as done in LT22 and in the present study.
A comparison of AUC scores suggests that higher accuracy in stability prediction can be achieved by densely sampling a single orbital element, rather than coarsely sampling multiple elements.
Furthermore, fine-grained sampling not only improves accuracy but also reduces the computational time required for classification.

\subsection{Summary and future work}
In this paper, we have carried out numerical simulations of hierarchical three-body systems in the mixed-region in order to explore a new method for determining the triple stability. 
Our main findings are as follows:
\begin{enumerate}
    \item Hierarchical triple systems in the mixed-region exhibit a wide distribution of survival times.
    The majority of systems in the mixed-region break apart within $10^5P_\mathrm{in}$ to $10^6P_\mathrm{in}$, but some systems exhibit significantly longer survival times, exceeding $10^9 P_\mathrm{in}$.
    Therefore, the current stability criterion based on $Q$ is inadequate for determining stability in the mixed-region.
    \item There is a obvious difference in the orbital evolution between stable and unstable triples in the first $10^3P_\mathrm{in}$.
    Stable triples exhibit more distinct peaks in the frequency space at the fundamental frequency and its integer multiples. 
    Additionally, stable ones tend to have smaller low-frequency components.
    There is a correlation between the triple’s instability and the stability of its orbit.
    It is remarkable that variations in orbital evolution already manifest within a mere $10^3 P_{\mathrm{in}}$ during the initial stages.
    \item The stability of a system can be predicted using the proportion of low-frequency components relative to the total power.
    This method facilitates the identification of a limited number of systems that attain stability within the range of $Q$ values situated near the vicinity of instability.
    Notably, our quantification relies solely on the low-frequency components.
    Combining this approach with other quantification methods is expected to enable predictions with even higher accuracy.
\end{enumerate}

Here, we present potential future works.
First, our findings are limited to coplanar cases, necessitating the verification of whether analogous discussions can be extended to inclination cases.
It is also necessary to investigate how the results change when orbital parameters such as mass and eccentricity are varied.
Second, it is plausible that incorporating eccentricity and other orbital parameters could yield a more precise prediction of stability.
For stable systems, other orbital elements should also exhibit nearly periodic variations.
Lastly, as a means to identify components that elude human visual perception, machine learning algorithms could be employed to analyze the FFT results and subsequently predict the stability of triples.




\begin{ack}
R. I. acknowledges support from JST SPRING, Grant Number JPMJSP2148.
Our numerical integration was mainly carried out on the general-purpose PC cluster at the Center for Computational Astrophysics, National Astronomical Observatory of Japan (NAOJ), and the CPU cluster ``\texttt{CPLAB}'' in Kobe University.
\end{ack}

\section*{Data availability}  

The data underlying this article will be shared on reasonable request to the corresponding author.


\bibliographystyle{aasjournal}
\bibliography{main}

\end{document}